\documentclass[aps,
               pra,
               twocolumn,
               amsmath,
               amssymb,
               superscriptaddress,
               reprint,
               10pt
              ]{revtex4-2}

\usepackage{url}
\usepackage{txfonts}

\usepackage{amsmath, amssymb}

\usepackage{pstricks}
\usepackage{graphicx}
\usepackage{tikz}
\usepackage[T1]{fontenc}
\usepackage{color}
\usepackage{chemformula}
\usepackage[colorlinks=true,breaklinks=true,allcolors=blue]{hyperref}

\usepackage{tabularx}
\usepackage{multirow}
\usepackage{array}

\usepackage{physics}
\usepackage{xparse}
\usepackage[version=4]{mhchem}
\usepackage[normalem]{ulem}
\usepackage{circledsteps}

\newcommand{\phis}{\varphi_\mathrm{S}}
\newcommand{\mf}{m_\mathrm{F}}
\newcommand{\Fperp}{F_{\perp}}
\newcommand{\Qyz}{Q_{yz}}

\makeatletter
\let\cat@comma@active\@empty
\makeatother

\begin{document}

%==============================================================================
%==============================================================================
\title{Experimental Observation of Topological and Non-Topological Bions}
%==============================================================================
\author{Alexander Schmutz}
\email{bions@matterwave.de}
\affiliation{Kirchhoff-Institut f\"{u}r Physik, Universit\"{a}t
Heidelberg, Im Neuenheimer Feld 227, 69120 Heidelberg, Germany}%

\author{Alexander Flamm}
\affiliation{Kirchhoff-Institut f\"{u}r Physik, Universit\"{a}t
Heidelberg, Im Neuenheimer Feld 227, 69120 Heidelberg, Germany}%

\author{Florian Schmitt}
\affiliation{Kirchhoff-Institut f\"{u}r Physik, Universit\"{a}t
Heidelberg, Im Neuenheimer Feld 227, 69120 Heidelberg, Germany}%

\author{Jan P. Ott}
\affiliation{Kirchhoff-Institut f\"{u}r Physik, Universit\"{a}t
Heidelberg, Im Neuenheimer Feld 227, 69120 Heidelberg, Germany}%

\author{Yannick Deller}
\affiliation{Kirchhoff-Institut f\"{u}r Physik, Universit\"{a}t
Heidelberg, Im Neuenheimer Feld 227, 69120 Heidelberg, Germany}%

\author{Thomas Gasenzer}
\affiliation{Kirchhoff-Institut f\"{u}r Physik, Universit\"{a}t
Heidelberg, Im Neuenheimer Feld 227, 69120 Heidelberg, Germany}%

\author{Panayotis G. Kevrekidis}
\affiliation{Department of Mathematics and Statistics, and Department of
Physics, University of Massachusetts Amherst, Amherst, 01003, MA, USA}%
\affiliation{Theoretical Sciences Visiting Program, Okinawa Institute of
Science and Technology Graduate University, Onna, 904-0495, Japan}

\author{Markus K. Oberthaler}
\affiliation{Kirchhoff-Institut f\"{u}r Physik, Universit\"{a}t
Heidelberg, Im Neuenheimer Feld 227, 69120 Heidelberg, Germany}%

\author{Helmut Strobel}
\affiliation{Kirchhoff-Institut f\"{u}r Physik, Universit\"{a}t
Heidelberg, Im Neuenheimer Feld 227, 69120 Heidelberg, Germany}%

\date{\today}

%==============================================================================
%Abstract
%==============================================================================
\begin{abstract}
Bions are spatially localized, time-periodic bound states of two domain walls in the integrable Landau–Lifshitz model. We experimentally realize topological and non-topological bions in a spin-1 Bose–Einstein condensate, in a regime where the dynamics, restricted to a suitable SU(2) subspace, is described by this effective model.
Deterministic phase imprinting enables systematic characterization of both bion branches, revealing opposite senses of rotation and frequency dependencies on amplitude and spatial extent in quantitative agreement with analytical predictions and full spin-1 simulations. Magnetic excitations outside this subspace modify the bion dynamics and limit their lifetimes, exposing the breakdown of the effective description.
\end{abstract}

%==============================================================================
%Introduction
%==============================================================================

\maketitle
\noindent Solitons play a foundational role in nonlinear physics and field theory \cite{Remoissenet1999,DauxoisPeyrard2006,ablowitz2011nonlinear}, spanning diverse systems such as nonlinear optics \cite{Kibler2010,LEDERER20081}, magnetic systems in condensed matter physics \cite{JosephsonJunction1992} and cold quantum gases ~\cite{djf,Abdullaev2005Bright},
among others.
In recent years, especially the study of solitons in multi-component quantum gases has intensified, with the experimental investigation of vector solitons~\cite{Bersano2018, Lannig2020}, magnetic solitons~\cite{Farolfi2020, Chai2020, Rabec2025}, sine-Gordon-like solitons~\cite{Deller2026}, Ferrodark solitons~\cite{Yu2021, Yu2022} and skyrmions~\cite{Leslie2011, Huh2025}.
Furthermore, collisional properties \cite{Bai2026, Deller2026} and breathing solutions \cite{Luo2020, Brochier2026} have been studied.\\
One of the theoretically most studied solitary wave
structures consists of the bions~\cite{KOSEVICH1990} of 
the Landau-Lifshitz equation (LLE) \cite{Landau:1935qbc} for magnetic systems.
Related waveforms have also been explored in the
sine-Gordon model~\cite{Cuevas-Maraver2014}.
Remarkably, the deterministic experimental generation of such bound states of pairs of domain walls has been challenging.\\
In this letter, we present the direct generation of both topological and non-topological types of easy-axis Landau-Lifshitz bions in a ferromagnetic spin-1 BEC via a phase imprinting protocol.
Our results constitute the prototypical experimental observation of the real time internal dynamics of bions.\\
The bions manifest as a rigid rotation of the spatial profiles of the spin-nematic vector in an SU(2) subspace of the spin-1 system~\cite{Kawaguchi2012}, leading to breathing behavior in the single components of the vector. The dynamics of the spin-nematic vector is well described by an LLE, considering constant total density. 
We characterize the frequency dependence of both types of bions on their amplitude and distance respectively, revealing a clear difference in direction of rotation, in accordance with analytical predictions and numerical simulations. Furthermore, we utilize spurious excitations generated by our imprint protocol acting as perturbations to probe the breakdown of the effective LLE (SU(2) subspace) description.\\

%==============================================================================
\begin{figure}
  \centering
    \includegraphics[width=0.5\textwidth]{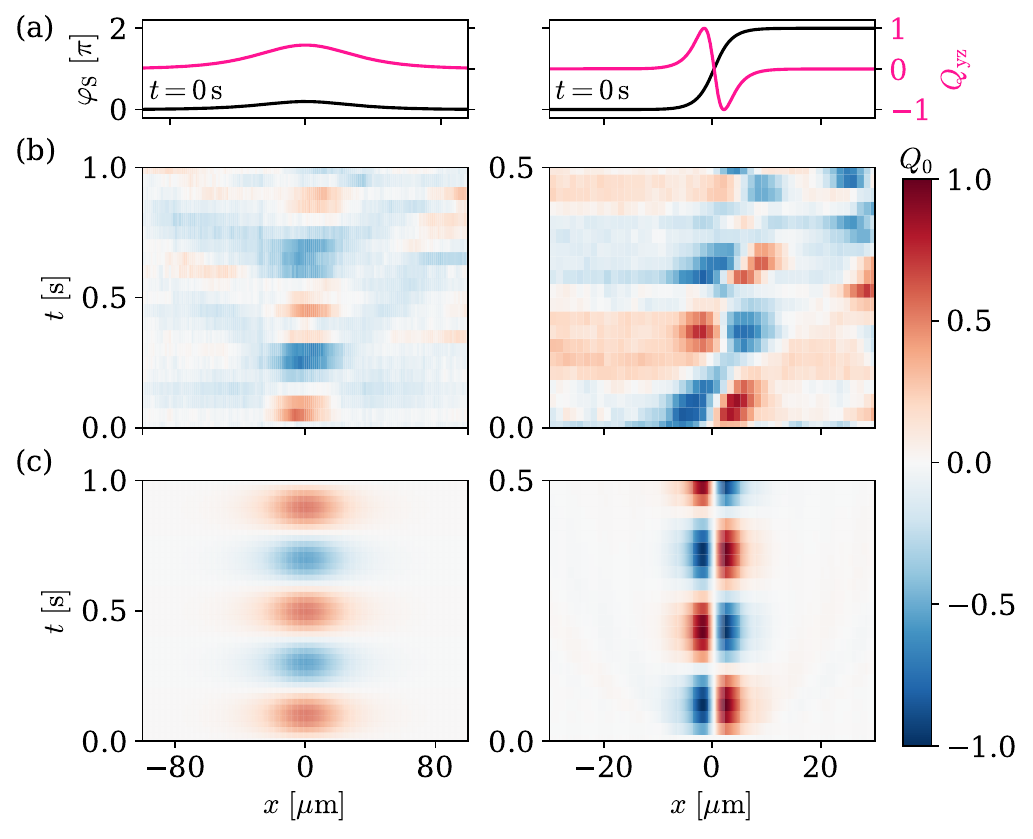}
  \caption{Observation of non-topological and topological bions. (a) Following initial imprints of the spinor phase $\phis$ (black), long-lived excitations emerge.
  The phase imprint on the left gives rise to a localized maximum in the quadrupole component $\Qyz = \sin{(\phis)}$ (pink), whereas the imprint shown on the right induces a phase difference of $2 \pi$ across the system and results in a double-lobe structure in $\Qyz$. (b) The subsequent time evolution shows oscillations in the relative side-mode population $Q_{\mathrm{0}}$. The non-topological bion (left) evolves into a single-lobe time-periodic structure, while the topological bion (right) shows a double-lobe periodic structure. The experimental data shows residual oscillations of the background, due to imperfections in the initial state preparation. (c) shows corresponding numerical simulations, for details see \ref{app:d}.}
    \label{fig:figure1}
\end{figure}
%==============================================================================

%==============================================================================
%Theoretical Description
%==============================================================================
\textit{Theoretical Description}---
The spinor BEC is described at the mean-field level by the Hamiltonian density
\begin{equation}\label{eq:equation1}
     H = \frac{\hbar^2 |\partial_x \boldsymbol{\psi}|^2}{2 M}+q\, \boldsymbol{\psi}^{\dagger} f_z^2 \boldsymbol{\psi} + \frac{c_0}{2} |\boldsymbol{\psi}^{\dagger} \boldsymbol{\psi}|^2 + \frac{c_1}{2} |\boldsymbol{\psi}^{\dagger} \boldsymbol{f} \boldsymbol{\psi}|^2\, .
 \end{equation}
Here $\boldsymbol{\psi} = (\psi_1, \psi_0, \psi_{-1})^T$ is the mean-field spinor wavefunction, whose components are given by $\psi_{\mf} = n_{\mf}^{1/2} e^{\mathrm{i}\varphi_{\mf}}$, where $n_{\mf}$ is the atomic density and $\varphi_{m_F}$ the respective phase of the Zeeman sublevels $m_F \in \left\{1,0,-1\right\}$ of the $F = 1$ hyperfine manifold. $\boldsymbol{f} = (f_x, f_y, f_z)^T$ are the spin-1 matrices \cite{Kawaguchi2012}, $c_0$ and $c_1$ are the density and spin interaction constants, respectively, $M$ is the mass of  the constituent particles, and $q$ the quadratic Zeeman shift, a relative energy shift of the $\mf =\pm1$ levels with respect to $\mf = 0$.\\
We consider a ferromagnetic spinor BEC with weak spin interaction, i.e., $|c_1| \ll c_0$, such that fluctuations in the total density $n = |\boldsymbol{\psi}^{\dagger} \boldsymbol{\psi}|$ are negligible, when preparing the system in the ground state of its trapping potential. We furthermore confine the discussion to states for which the local magnetization $F_z(x) = n_1(x) - n_{-1}(x)$ along the quantization axis vanishes. The system's dynamics is then well captured in a closed SU(2) subspace given by the so called spin-nematic subspace~\cite{gerving2012non}. This subspace is spanned by the three operators $F_x, \Qyz, Q_0$, which we collect into the spin-nematic vector  $\boldsymbol{m}$ as 
\begin{equation}
    \boldsymbol{m} = (F_x, \Qyz,Q_{\mathrm{0}}) \, .
    \label{eq:equation2}
\end{equation}
For vanishing quadratic Zeeman shift $q=0$ , the equation of motion for $\boldsymbol{m}$ assumes a simple form and leads to an easy-axis LLE  given by
\begin{equation}
    \partial_T \boldsymbol{m} = (\partial_X^2\boldsymbol{m}+m_1\boldsymbol{e}_1)\times\boldsymbol{m} \, ,
    \label{eq:equation3}
\end{equation}
where the anisotropy favors alignment along \(\boldsymbol{m}=\pm\boldsymbol{e}_1\) as shown in~\cite{landau1984electrodynamics}.
Here, $\boldsymbol{e}_1=(1,0,0)$ denotes the unit vector along the first axis of the spin-nematic space.
The entries of the vector $\boldsymbol{m}$ can be obtained by evaluating $\psi_{m'_F}^{*} o_{m'_F m_F} \psi_{m_F} / n$ using the corresponding matrices listed in \ref{app:c}.
The dimensionless time and spatial coordinates $T = 2t/t_s$ and $X = (x-x_0)/\ell$ are defined by a spatial offset $x_0$, the characteristic time $t_s = \hbar/(n \lvert c_1\rvert) $ and the characteristic length scale $\ell = \xi_\mathrm{s}/\sqrt{2}$ with $\xi_\mathrm{s} = \hbar / \sqrt{2 M n \lvert c_1 \rvert}$. 
While similar mappings of multi-component systems onto LLEs have been discussed previously \cite{Li2005,Brochier2026}, a corresponding description of spin-1 dynamics in the spin-nematic manifold has not been obtained. Here, we establish such a mapping, with a detailed derivation given in Ref.~\cite{Schmitt2026}.\\

%==============================================================================
%Experimental system
%==============================================================================
\textit{Experimental system}---
We prepare a quasi one-dimensional (1D) spin-1 BEC of $\sim 10^5$ \ch{^{87}Rb} atoms in a harmonic trap, with trapping frequencies $(\omega_\parallel,\omega_\perp) \approx 2\pi\times(1.6,170)\ \mathrm{Hz}$, in a weak magnetic offset field of $\sim 0.9 \, \mathrm{G}$. The offset field gives rise to a quadratic Zeemann shift $q/h \sim 58 \, \mathrm{Hz}$, which we compensate with off-resonant microwave-dressing to $q/h = (0.0\pm0.2) \, \mathrm{Hz}$. We furthermore work for all experiments reported here in the sufficiently homogeneous region of the harmonic trap.\\
We prepare the system in a fully elongated transversal spin state which is a ground state for $q/h=0$ and ferromagnetic spin interactions, with a combination of radiofrequency and microwave pulses at the start of each experimental run.\\
In order to generate the bions we make use of a local spin manipulation protocol described in~\cite{Deller2026}, to locally vary $\Qyz$ via a magnetic field generated by the vector Stark shift of the local laser beam.
The technique allows to imprint a sum of spatial Gaussian profiles in $\Qyz$ with a resolution of $\sim4 \ \mathrm{\mu m}$ and variable amplitude. We can approximate a desired imprint function by varying the illumination time and successive distances of the individual imprints. See~\ref{app:a1} for further details.\\

%==============================================================================
\begin{figure}
  \centering
    \includegraphics[width=0.5\textwidth]{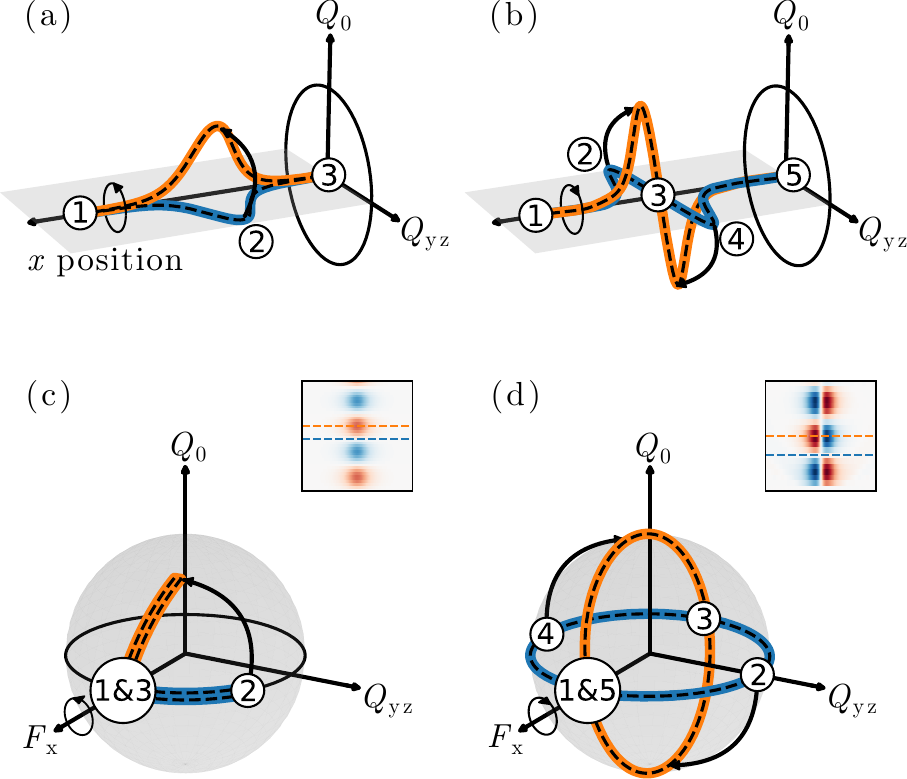}
  \caption{Conceptual representation of the bions' dynamics in the spin-nematic subspace. Numerical spatial profiles of (a) the non-topological and (b) topological bion at two representative evolution times (blue and orange, see dashed lines in the insets) in the observables $Q_{\mathrm{0}}$ and $\Qyz$, together with (c,d) their corresponding trajectories on the spin-nematic sphere spanned by $F_x$, $\Qyz$, and $Q_{\mathrm{0}}$. For both excitations, the oscillatory dynamics in $Q_{\mathrm{0}}$ and $\Qyz$ is a result of a rigid rotation of the spin-nematic vector defined in Eq.~\eqref{eq:equation2} around the $F_x$-axis. The orientation of rotation shows opposite sign as indicated by the arrows. The non-topological bion spans an arc (\Circled{1} to \Circled{3}) whose angular extent determines its amplitude $A$, whereas the topological bion traces a closed $2\pi$ loop (\Circled{1} to \Circled{5}), cf. black graphs in Fig.~\ref{fig:figure1}a.}
    \label{fig:figure2}
\end{figure}
%==============================================================================

%==============================================================================
\begin{figure*}
  \centering
    \includegraphics[width=1\linewidth]{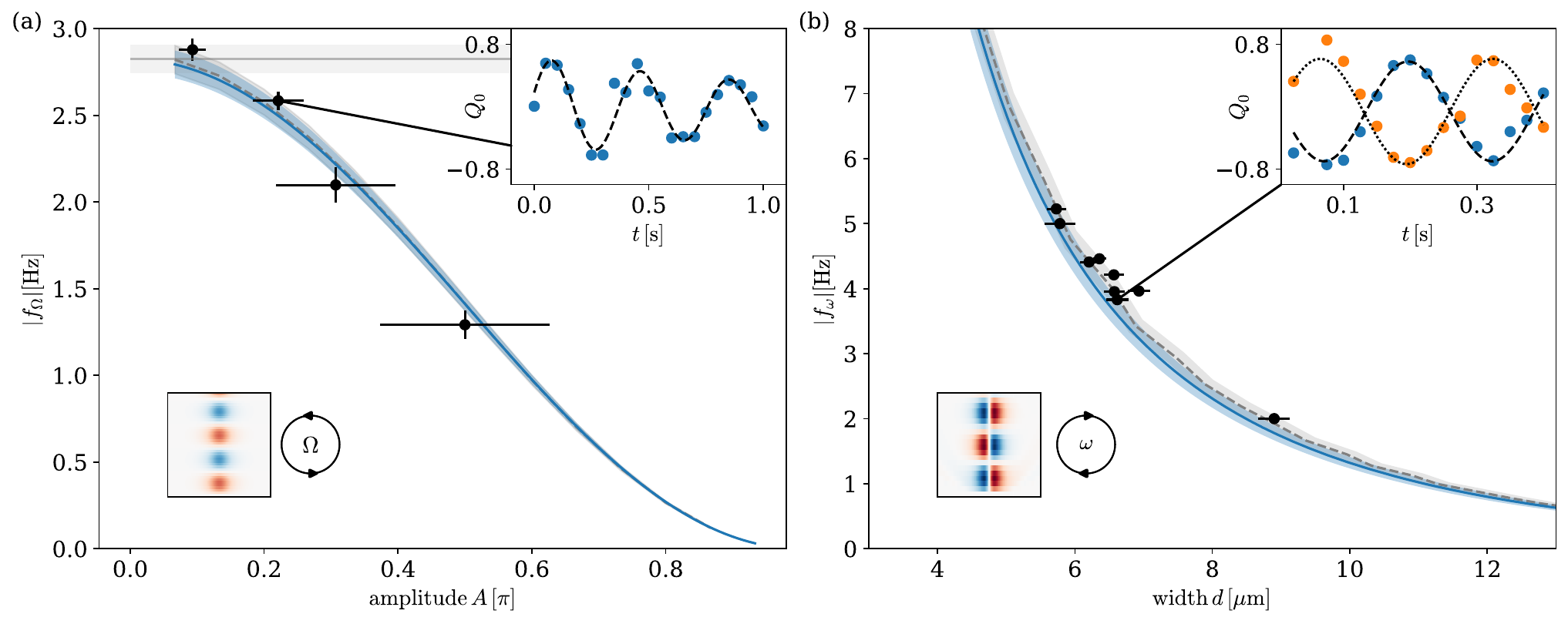}
  \caption{Frequency dependence of non-topological and topological bion.
(a) The non-topological bions frequency $f_\Omega = \Omega /(\pi t_s)$  (black dots) for varying spinor phase amplitude $A$ is extracted via a damped oscillation fit (dashed line) to
$Q_{0}$. An exemplary fit is shown in the upper inset resulting in $f_\Omega = 2.58(5)\,\mathrm{Hz}$, amplitude $0.64(11)$ and a characteristic damping time of $ 1.5(8)\,\mathrm{s}$.
The non-topological bion approaches an upper bound in
the low-amplitude limit given by the experimentally extracted background oscillation (grey upper line). (b) The topological bion frequency $f_\omega = \omega /(\pi t_s)$ (black dots) for varying width $d$ is obtained by extracting $Q_{\mathrm{0}}$ at the respective maxima (orange and blue dots) and subsequently employing an oscillation fit (dashed and dotted lines), as depicted in the inset. The exemplary fit retrieves $f_\omega = \,3.91(5)\mathrm{Hz}$ and an amplitude of $0.66(2)$ for a width $d = 6.93(16)\,\mathrm{\mu m}$.
The data for both bions show quantitative agreement with analytical predictions (blue line) and numerical simulations (grey line) within errors.
Experimental error bars represent the fit uncertainty in frequency, amplitude and width. The numerical
error band in both cases reflects the experimental uncertainties of the parameters used for rescaling the analytical prediction and numerical data to experimental dimensions.
}
    \label{fig:figure3}
\end{figure*}
%==============================================================================

%==============================================================================
%Observation of bions
%==============================================================================
\textit{Observation of bions}---
The easy-axis LLE is known to possess bion solutions that feature a positive- and negative frequency branch~\cite{Kosevich1990Magnetic}. 
We will distinguish between the two branches by using the terminology of non-topological vs.~topological bions as shown in Fig.~\ref{fig:figure1}a. Here the topological character is determined by the difference in the spinor phase $\phis = \arctan{\left(m_2/m_1\right)}$ across the system, keeping the nomenclature in accordance with the foundational literature \cite{KOSEVICH1990}. In the spin-nematic parameterization discussed here both branches show ($\pi/2$)-out-of-phase oscillations in the observables $Q_{\mathrm{0}}$, $\Qyz$, which are given by 
\begin{align}
    % \left.
    \Qyz^{\mathrm{nT}}(X,T)  &= \frac{2 \sqrt{\Omega(1-\Omega)}\,{\cosh{(\sqrt{1-\Omega}X)}}}{1+\Omega\sinh^2{(\sqrt{1-\Omega}X)}}\cos{(\Omega T)}
     \label{eq:equation4}
    \\
    \Qyz^{\mathrm{T}}(X,T) &= \frac{2 \sqrt{\omega(1+\omega)}\sinh{(\sqrt{1+\omega}X)}}{1+\omega\cosh^2{(\sqrt{1+\omega}X)}}\, \cos{(-\omega T)} 
     \label{eq:equation5}
\end{align}
for non-moving bions, where nT and T indicate non-topological and topological respectively. In principle, the LLE also admits solutions in the form of moving bions, however, these lie beyond the scope of the present work.
Here, $\Omega = \pi f_\Omega t_s$ and $\omega= \pi f_\omega t_s$ are the corresponding dimensionless frequencies, for which $0<\Omega < 1$ and $\omega>0$ holds.
The bions show an opposite orientation of rotation, as can be seen from the sign of the frequency argument. The full vector solutions including $F_x(X,T)$ and $Q_0(X,T)$ are given in \ref{app:c}.\\
We imprint profiles close to \eqref{eq:equation4} and \eqref{eq:equation5} for $T = 0$, limited by the resolution of our imprint. 
Following the imprint we observe the formation of stable objects in the relative side-mode population $Q_{\mathrm{0}} = n_0-(n_1+n_{-1})$, see Fig.~\ref{fig:figure1}b. Numerical spin-1 simulations confirm the stability of the emerging objects (see Fig.~\ref{fig:figure1}c) in the experimentally shown time window.
The emerging non-topological bion exhibits a single-lobe oscillation in $Q_{\mathrm{0}}$ while the topological bion shows a pair of counter oscillating lobes. The latter is reminiscent of the Flemish string in the anti-ferromagnetic regime which also oscillates in a spin-nematic subspace \cite{FlemishStringsUeda}. The observable $Q_{\mathrm{0}}$ is used for all quantitative characterizations as it can be accessed robustly in the experiment.\\
Compared to the numerical results, the background of the experimental data shows residual amplitude oscillations that can be distinguished from the bions in the data analysis and show no effect on their oscillation behavior.

%==============================================================================
%Bion dynamics on the spin-nematic sphere
%==============================================================================
\textit{Bion dynamics on the spin-nematic sphere}---
According to the vector parameterization \eqref{eq:equation2} and equation of motion \eqref{eq:equation3} in the spin-nematic sub-space the bions' dynamics is given by a rigid rotation around $F_x$ on the spin-nematic sphere. 
Figs.~\ref{fig:figure2}a,b 
show the numerical spatial profiles of $\Qyz$ and $Q_{0}$ at two representative time steps (see dashed lines in insets), alongside the corresponding trajectory on the spin-nematic sphere, as depicted in panels c and d.\\
The spatial profile maps onto an arc on the sphere, and its oscillation in
$Q_{\mathrm{0}}$ and $\Qyz$ is equivalent to the arc's rotation. The two bions differ in the length of this arc. The non-topological bion's extent is determined by its imprinted amplitude, while the topological bion always traces a full $2\pi$ circle around the sphere.  As $F_x$ constitutes the rotation axis, the transversal spin length is preserved. As indicated in Fig.~\ref{fig:figure2}, the non-topological and topological bions exhibit opposite orientation of rotation, in agreement with the analytical LLE predictions for the two branches.\\
For states confined close to the equator of the spin-nematic sphere, as is e.g.~the case for low-amplitude non-topological bions, an effective description of their dynamics in the framework of a  sine-Gordon equation can be recovered from the LLE \cite{Schmitt2026, Faddeev2007}, which is in accordance with the results presented in \cite{Siovitz2025, Deller2026}.\\

\textit{Characterization of Oscillatory Dynamics}---
The imprint resolution and experimental stability allow us to systematically characterize the bions' oscillatory behavior quantitatively. 
The non-topological bion features a distinct interdependence between the oscillation frequency $\Omega$ and the extent of the arc spanned on the spin-nematic sphere as discussed in Fig.~\ref{fig:figure2}c. We parametrize the arc length as $A=\arctan\!\left[\Qyz(0,0)/F_x(0,0)\right]$
in order to compare analytical predictions with experimental and numerical spin-1 results. 
We find the dependence
\begin{equation}
    \Omega = \cos^2{(A/2)} \, 
    \label{eq:equation6}
\end{equation}
between amplitude and frequency by evaluating Eq.~\eqref{eq:equation4} with the aforementioned definition of the amplitude $A$, see \ref{app:c1} for details. We vary the imprint amplitude following Eq.~\eqref{eq:equation4} and extract the non-topological bion's frequency as follows. \\
We employ a Gaussian fit to the spatial profiles in $Q_{\mathrm{0}}$, extract the amplitude at each instance of time and subsequently fit a damped oscillation to the time series and retrieve the frequency $f_\Omega = \Omega/(\pi t_\mathrm{s})$, as depicted in the upper inset of Fig.~\ref{fig:figure3}a. As we do not exceed arc lengths of $\pi/2$ in the experiments, we can uniquely extract the amplitude $A$ from the measured $Q_0$ data via $A = \arcsin{\left[ Q_{0}(0,\pi /(2 \Omega)) \right]}$ via the amplitude of the oscillation. In the low-amplitude limit the non-topological bion approaches a small amplitude background oscillation (see Fig.~\ref{fig:figure3} grey upper line). The results are in quantitative agreement with numerical spin-1 simulations within the error bars and with the analytical prediction of Eq.~\eqref{eq:equation6}. In the limit $A\to0$ the object merges into a breather solution of the effective sine-Gordon equation.\\
Concerning the topological bion the frequency depends on the separation of the two lobes depicted in Fig.~\ref{fig:figure2}b. One can calculate the frequency dependence on their separation by determining the positions of the extrema of Eq.~\eqref{eq:equation5} and taking their difference, which leads to the prediction
\begin{equation}
        d = \frac{\ell}{\sqrt{1+\omega}}\mathrm{arcosh}\bigg({3+\frac{2}{\omega}}\bigg)
    \label{eq:equation7} \, ,
\end{equation}
see \ref{app:c2} for details.
In the spin-1 system, one can extract the width $d$ most robustly from the observable $F_x$, where the local extrema in $\Qyz$ lead to a zero crossing in the transversal spin length. Denoting the zeros by $x^{\pm}$, we extract the width as $d = x^+ - x^-$. In order to extract the frequency dependence on the width of the topological bion we employ a simultaneous readout of the transversal spin length and $Q_{\mathrm{0}}$ \cite{Kunkel2019} and extract the time evolution of the counter oscillating lobes at the positions of vanishing $F_x$, corresponding to the respective maxima in $Q_{\mathrm{0}}$ (see \ref{app:b1}). As shown in Fig.~\ref{fig:figure3}b, we use a sinusoidal fit, as is depicted in the inset, finding $f_\omega = \omega/(\pi t_s)$ to increase with decreasing width $d$ in quantitative agreement with simulations and the analytical prediction of Eq.~\eqref{eq:equation7} within error bars.\\

%==============================================================================
\begin{figure}
    \centering
    \includegraphics[width=1\linewidth]{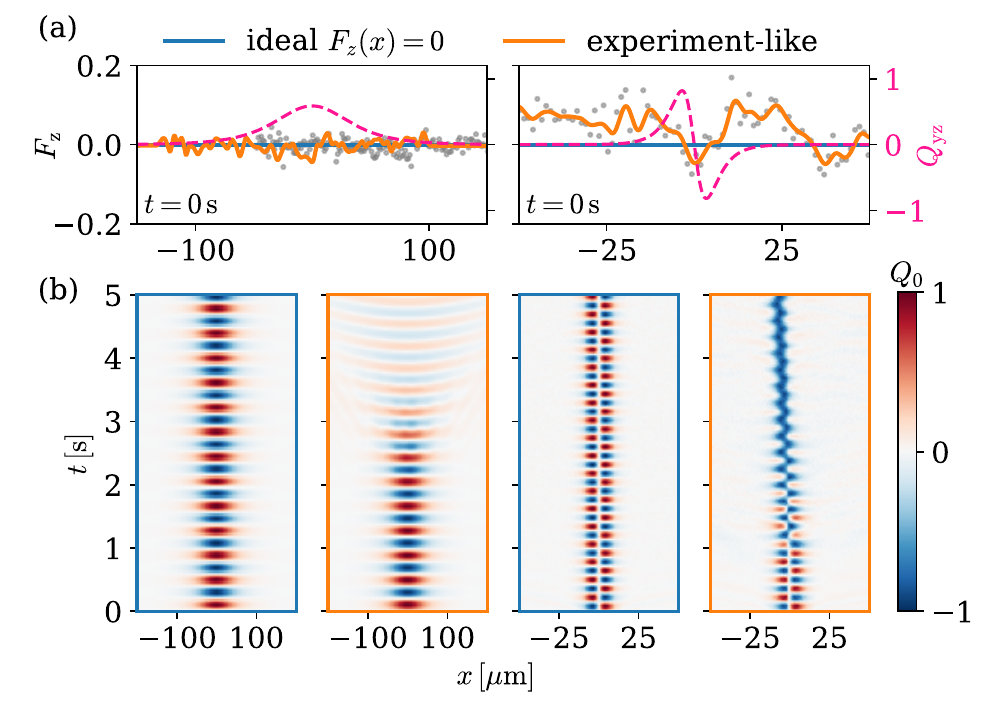}
    \caption{Probing effects beyond the effective LLE description by utilizing spurious $F_z$ excitations. a) The numerical initial conditions in $F_z$ for an ideal preparation with vanishing magnetization (blue) and the experimental initial condition (grey dots), spatially smoothed (orange) at the position of the imprint indicated by the $\Qyz$ profile (pink). b) Time series of non-topological (left) and topological bion (right) with (orange frame) and without (blue frame) spurious $F_z$ effects. The simulations show a significant reduction in lifetime of both objects on the same order of magnitude as in the experiment when introducing $F_z$ excitations. The topological bion furthermore decays into a stable excitation in $Q_0$, as is visible on the right.}
    \label{fig:figure4}
\end{figure}
%==============================================================================

%==============================================================================
%Observable Timescales
%==============================================================================
\textit{Observable Timescales---}
The validity of the LLE hinges on the assumption of the spin-nematic subspace being a closed SU(2) subspace. However, in the full spin-1 system, states can leave the subspace by acquiring an expectation value in another spin observable, leading to more intricate dynamics that go beyond the predictions of the integrable LLE. 
In the experiment the chosen tune-out wavelength of the imprinting laser leads to the cancellation of the light shifts for $F = 1, m_F = 0$ \cite{Schmidt_2016}, whereas the $m_F = \pm1$ levels are subject to a non-zero vector Stark shift. The non-constant intensity profile leads to a relative spatial displacement between the side-modes, thereby creating $F_z$ excitations over the spatial extent of the imprint. Interestingly, these excitations provide a direct probe of the limitations of the effective LLE description beyond the timescales shown in Fig.~\ref{fig:figure1}.
Fig.~\ref{fig:figure4}a shows the experimentally observed local $F_z$ excitations for both bions directly after the imprint. The imprint amplitude is lower for the non-topological bion (left) than for the topological bion (right) displayed here. We observe stronger effects with higher amplitude imprints, corresponding to higher intensities in the local beam.  
In order to investigate the effect of the spurious excitations, we add the spatially smoothed experimental initial condition in $F_z$ to the numerical simulation. 
Fig.~\ref{fig:figure4}b reveals a decrease in lifetime on the same order of magnitude as in the experimental data. We observe a damped oscillation of the non-topological bion and an amplitude $<1$ in $Q_{\mathrm{0}}$ of the topological bion, which is in accordance with experimental observations shown in the upper insets of Fig.~\ref{fig:figure3}.
The experimental $F_z$ excitations are weaker for small-amplitude bions, nonetheless, the evaluation of the frequency dependence is limited by a decrease in signal to noise ratio over the evolution time, making the quantitative data analysis unreliable for $t > 1s$. However, we can still identify localized profiles in single shots for up to $t=2s$.\\
The additional inclusion of $F_z$ into the dynamics leads to a richer behavior of both bions' dynamics. While the non-topological bion decays into a lower amplitude global oscillation, the topological bion decays into a localized excitation, which manifests as a localized bump in $Q_0$, raising the question whether it retains its topological character in other spin observables. While beyond the scope of the present work, this is
an interesting topic for further investigation.\\

%=============================================================================
%Discussion and outlook
%=============================================================================
\textit{Discussion and outlook---}
We have established the existence of a non-topological and a topological bion in spin-1 BECs. The systematic frequency dependence on their spatial profiles is in accordance with analytical predictions. As expected from the LLE their dynamics is captured by a rigid rotation in the spin-nematic subspace and we found that spurious $F_z$ excitations lead to more intricate dynamics going beyond the LLE predictions and modifying the shape of both excitations.\\
Experimentally, a natural next step is to probe bion formation by domain wall collisions, thereby characterizing the excitation as a genuine bound state. The formation and existence of 
moving bion states and their morphing into
higher-order soliton states as a result of bion collisions is furthermore an interesting avenue. Bion states in higher spatial dimensions and their 
potential transverse stability also promises to be a topic of intriguing topological
consequences~\cite{Ruth2018}.
The experimental control of the topological bion enables the investigation of its impact onto the far-from-equilibrium universal dynamics observed in quenched ferromagnetic spinor condensates in a similar vein to the numerical investigation in \cite{FlemishStringsUeda} for the case of an antiferromagnetic spinor BEC. \\

%==============================================================================
%Acknowledgments
%==============================================================================
\textit{Acknowledgments---}
The authors thank Raphael Schäfer for numerical work during the beginning of the project.
They acknowledge support 
by the Deutsche Forschungsgemeinschaft (DFG, German Research Foundation), through 
SFB 1225 ISOQUANT (Project-ID 273811115), 
grant GA677/10-1, 
and under Germany's Excellence Strategy -- EXC 2181/1 -- 390900948 (the Heidelberg STRUCTURES Excellence Cluster), 
and by the state of Baden-W{\"u}rttemberg through bwHPC and the DFG through 
grants 
INST 35/1503-1 FUGG, INST 35/1597-1 FUGG, and INST 40/575-1 FUGG
(SDS, Helix, and JUSTUS2 clusters).
This material is partially based upon work supported by the U.S. National Science Foundation under the award PHY-2408988 (PGK). 
This work was also supported by the Simons Foundation
[SFI-MPS-SFM-00011048, P.G.K.].\\
In developing the theoretical description of the experimentally observed spin-nematic dynamics, we used the ChatGPT model GPT-5.6 Sol as a research aid to explore possible effective theories. The model suggested a connection to the Landau–Lifshitz framework, which led to an effective description of the dynamics beyond the previously considered sine-Gordon model. The resulting theoretical connection, its analytical treatment, and its comparison with numerical simulations and experimental observations were subsequently derived and validated by the authors.\\

%=============================================================
%APPEDIX
%=============================================================
\clearpage
\begin{appendix}

\onecolumngrid
\vspace{1cm}
\begin{center}
\textbf{APPENDIX}
\end{center}
\setcounter{equation}{0}
\setcounter{table}{0}
\makeatletter

In this Appendix, we provide further details on the experimental system, data analysis,  analytical solutions of the Landau-Lifshitz equation and numerical simulations of the spin-1 system.

%==============================================================================
%==============================================================================
\section{Experimental system}
At the start of every experimental sequence reported here, we prepare the BEC close to an easy-plane ground state of $q/h=0$, which corresponds to an elongated transversal spin  with $|\Fperp| \approx 1$. For this, all atoms are first transferred to $\ket{F=1,\mf=0}$, followed by a $(\pi/4)$ radio frequency (RF) pulse which prepares a state with $\Qyz = 1$. In order to align the spin onto the $F_x$-axis, we null the global spinor phase $\phis = (\varphi_1 + \varphi_{-1})/2 - \varphi_0$. This is realized by two subsequent MW $\pi$-pulses between $\ket{F=1,\mf=0}$ and $\ket{F=2,\mf=0}$ with the relative phase between the first and second pulse determining the change in spinor phase.
%==============================================================================
\subsection{Local manipulation of the spinor phase}
\label{app:a1}
\begin{figure}
    \centering
    \includegraphics[width=0.5\linewidth]{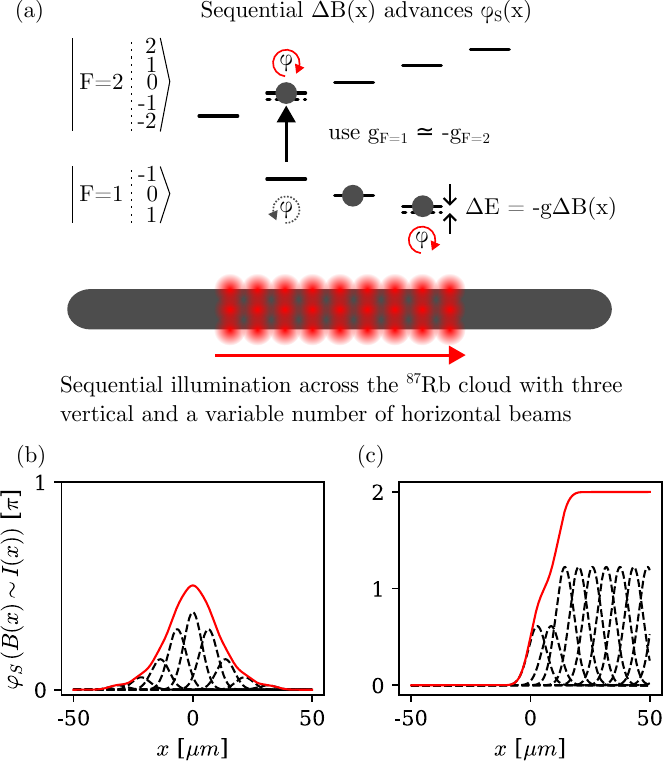}
    \caption{Localized spinor phase imprint protocol. (a) We advance the spinor phase by holding the atoms in a configuration, where the phases of the side modes advance in the same direction with respect to $\mf=0$. A steerable laser beam at the tune-out wavelength induces an additional local magnetic field $\Delta B(x)$ proportional to the beam intensity via the vector Stark shift. By varying the beam duration, we can imprint different initial profiles, as shown in (b) close to a non-topological bion, here in the case of $A = \pi/2$, and (c) close to a topological bion with a width of $d \sim 9\,\mathrm{\mu m}$.}
    \label{fig:figure5}
\end{figure}

The spinor phase imprinting protocol was first discussed in \cite{Deller2026}, where a detailed discussion of the sequence is given. In the following, we summarize the main idea.\\
In order to locally manipulate the spinor phase $\phis$, we advance the sum of the phases of the side modes $m_F = \pm1$ relative to the phase of the zero-mode $m_F = 0$. During the imprint we hold the atoms in a configuration of hyperfine states as shown in Fig.~\ref{fig:figure5}a, such that the sum of the side mode phases advances proportional to the linear Zeeman shift, as $g_{F=1} \approx - g_{F=2}$ for the electronic ground state. In order to locally manipulate the spinor phase, we sequentially illuminate the cloud with a steerable Gaussian laser beam with a waist of $w \sim 7 \,\mathrm{\mu m}$ (this corresponds to the radius of the FWHM, $r_{\mathrm{FWHM}} = w \sqrt{\ln{2}/2} \sim 4 \,\mathrm{\mu m}$), which we operate at the tune-out wavelength of $790\,\mathrm{nm}$ to minimize spurious excitations of the total density~\cite{Schmidt_2016}.
The circularly polarized laser beam generates a magnetic field proportional to the intensity of the laser via the vector Stark shift, which in turn locally advances the spinor phase. A combination of multiple Gaussian beams with matched spacings and individual illumination times leads to a spinor phase profile close to the targeted profile of both non-topological bion Eq.\eqref{eq:equation4} and topological bion Eq.\eqref{eq:equation5}, see Fig.~\ref{fig:figure5}b,c respectively.

%==============================================================================
%==============================================================================
\section{Additional details on the data analysis and bion profiles}
In the following section, we give additional details on the data analysis employed for the extraction of the frequency dependence of the topological bion's width and compare the analytically predicted profiles to experimental realizations.

%==============================================================================
\subsection{Topological bion analysis}
\label{app:b1}
\begin{figure}
    \centering
    \includegraphics[width=0.5\linewidth]{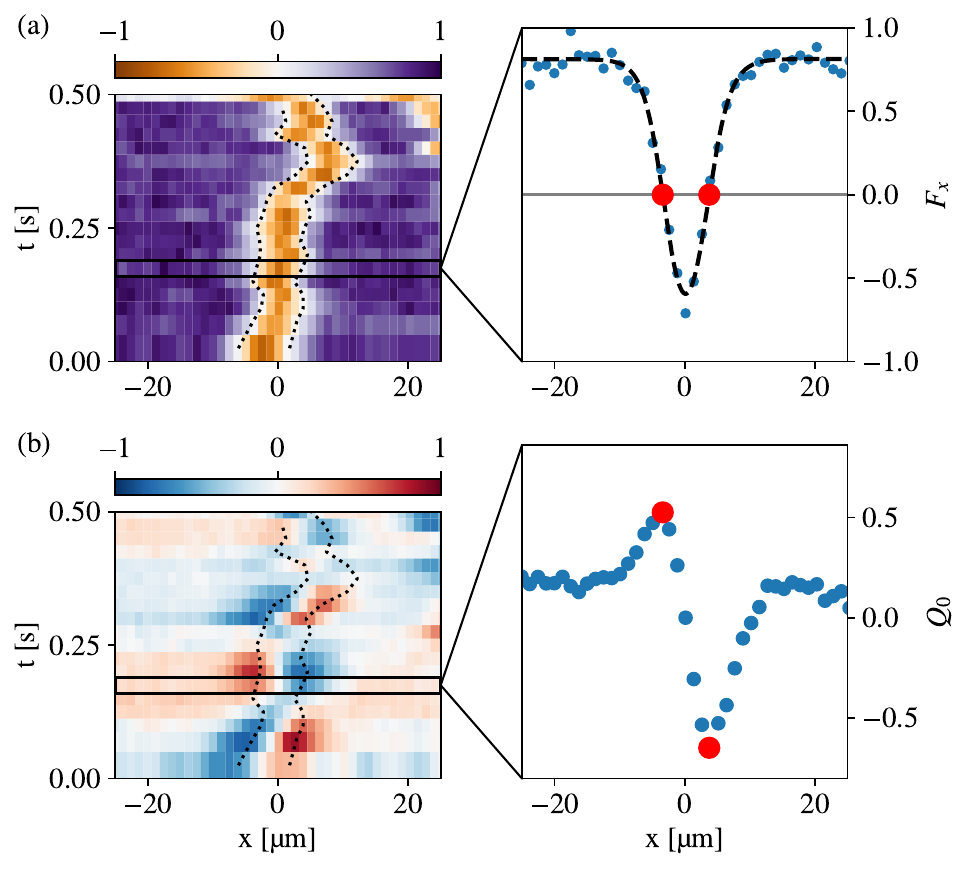}
    \caption{Exemplary data analysis for extracting the topological bions frequency dependence on its width. The depicted data shows means over 3-5 single shots.  (a) shows the stationary transversal spin data of $F_x$ up to $0.5\,\mathrm{s}$ while (b) shows the corresponding time dependent evolution of $Q_0$. Note, that both observables were extracted in the same measurement. The spatial profiles on the right show the experimental data (blue dots) with a fit of Eq.~\eqref{app:equationB1} (dashed line). We extract the central positions (red dots) from the zero crossings of $F_x$, resulting in a width of $6.93(16)\,\mathrm{\mu m}$, and evaluate $Q_0$ at those positions in order to obtain the time evolution of the $Q_0$ amplitudes shown in the inset in Fig.~\ref{fig:figure3}b. We perform a sine fit on each slice and mean over the results to obtain: $f_ {\omega} = 3.91(5)\,\mathrm{Hz}$ and amplitude $a = 0.66(2)$ .}
    \label{fig:figure6}
\end{figure}

To extract the topological bion frequency, we employed a POVM scheme that enables the simultaneous readout of the transverse spin, $F_\perp = F_x + iF_y$, and $Q_0$~\cite{Kunkel2019}. As discussed in the main text, the topological bion's maxima of the double-lobe structure coincide with vanishing transversal spin $F_x$. While the transversal spin profile remains stationary, see Fig.~\ref{fig:figure6}a, $Q_0$ undergoes a damped periodic time evolution, see Fig.~\ref{fig:figure6}b. We extract the position of vanishing $F_x$ by employing a fit given by
\begin{equation}
    F_x = \sum_{i = 1}^2 A_i\tanh{[(x-x_i)/l_i]}
    \label{app:equationB1}
\end{equation}
for each evolution time, as depicted in the profile cut of~Fig.~\ref{fig:figure6}a. The fit function is motivated by two isolated domain walls for the transversal spin which is closely related to Eq.~\eqref{eq:equation4}.
We then evaluate $Q_0$ at $x_i$ (red dots in spatial profile plots in Fig.~\ref{fig:figure6}a, b) which gives rise to the oscillations depicted in the inset of Fig.~\ref{fig:figure3}b in the main text. The width of the topological bion is defined as $d=x_2-x_1$

%==============================================================================
%==============================================================================

%==============================================================================
\subsection{Comparison of the bion profiles to experimental realizations}
\label{app:b2}
%==============================================================================
\begin{figure}
    \centering
    \includegraphics[width=0.5\linewidth]{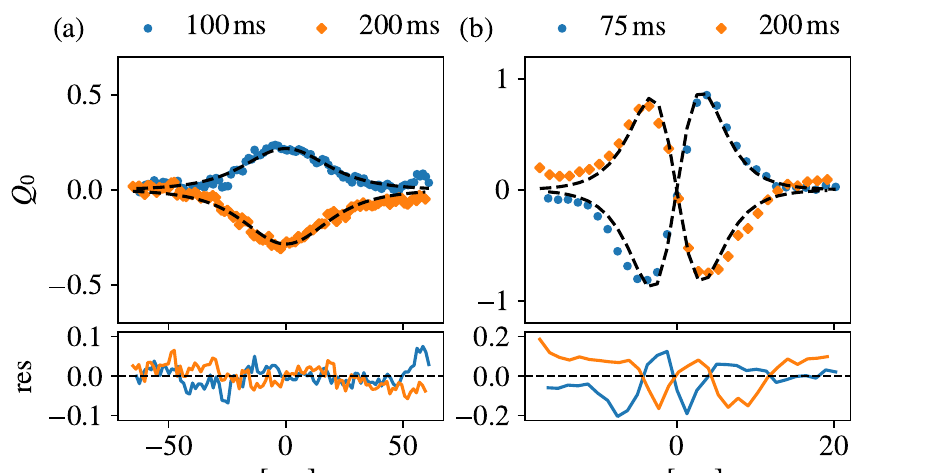}
    \caption{Comparison of the analytical profiles predicted by the LLE to the experimentally extracted spatial profiles. (a) shows experimental profiles of the non-topological bion at two representative time steps (blue dots \& orange diamonds). We fit Eq.~\eqref{app:equationC5} to the experimental profiles (black dashed lines) by fixing $\Omega$ using the extracted frequencies of Fig.~\ref{fig:figure3} and the characteristic time $t_s$ and length scale $ \xi_s$ given in Tab.~\ref{tab:table1}. We treat the time $t$, position $x_0$, and an additional offset $c$ as free parameters. The offset accounts for residual background oscillations and is subtracted from the data and fitted profile for visualization. (b) Experimental profiles at two representative time steps of the topological bion, together with fits to the analytical profiles given by \eqref{app:equationC11}. Here, only the position $x_0$ and time $t$ are treated as free parameters. The lower panels show the residuals of the fits with the experimental profiles.}

    \label{fig:figure7}
\end{figure}
%==============================================================================
As discussed in the main text, the bion's characteristic frequency dependence and rotational dynamics are in quantitative agreement with the analytical easy-axis LLE. While the analysis presented above is sufficient to extract and compare the characteristic frequencies, it does not rely on the detailed spatial profiles of the analytical solutions. We therefore provide a separate comparison of the analytically predicted spatial profiles with the experimental results for both the non-topological and the topological bion.
For this comparison, we use the results of $f_\Omega$ and $f_\omega$, (see Fig.~\ref{fig:figure3}), and the characteristic time scale $t_s$ and length scale $\xi_s$ as shown in Tab.~\ref{tab:table1} to determine $\Omega=\pi f_\Omega t_s$ and $\omega=\pi f_\omega t_s$ respectively. Fig.~\ref{fig:figure7} shows the resulting fitted spatial profiles of Eqs.~\eqref{app:equationC5} and \eqref{app:equationC11} (black dashed lines) at two representative time steps (blue dots and orange diamonds) each for the non-topological bion (see a) and the topological bion (see b). The fit uses $x_0$ and $t$ as free parameters, with $t$ setting the global amplitude, and $c$ as an additional offset.
 The offset accounts for non-vanishing oscillatory dynamics of the transversal spin background. The experimental profiles are well described by the analytical prediction, as shown by the residuals in the lower panels, further corroborating the results presented in the main text.

%==============================================================================
\subsection{Conversion of numerical units to experimental units from background dynamics}
\label{app:b3}
\begin{figure}
    \centering
    \includegraphics[width=0.5\linewidth]{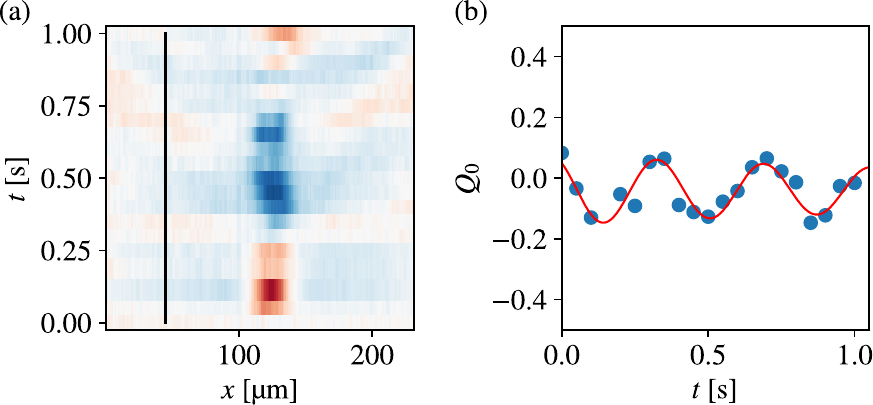}
    \caption{Extraction of the reference value for the spin background oscillation frequency. In (a) we indicate the spatial slice used to evaluate the backgrounds oscillation dynamics (black line), showing that it is well separated from the localized bion excitation ($A_B \sim 0.5$) in the center. In (b), the time evolution of $Q_0$ (blue dots), and a damped sine fit to the data (red line) are shown, resulting in a background frequency of $f_{b, exp} = 2.74(7) \, \mathrm{Hz}$ and characteristic damping time $\tau = 3(3)\,\mathrm{s}$.}
    \label{fig:figure8}
\end{figure}

In the following we discuss the data analysis pertaining to the reference (grey bar) in Fig.~\ref{fig:figure3} in the main text and furthermore explain the conversion procedure between experimental and numerical units.
As mentioned above, in addition to the localized breathing excitation, the transversal spin background, and therefore also $Q_0$, undergo oscillations due to non-perfect preparation of the initial condition. Fig.~\ref{fig:figure8} shows the extraction of the background oscillation for a single spatial cut. We generate the reference point (grey bar) shown in Fig.~\ref{fig:figure3} by taking a mean over 40 spatial cuts, in order to capture the non-constant total densities effect on the background dynamics. In order to extract the frequency $f_{b, exp}$, we fit a damped sine (red line) to the time series obtained from the mean of the spatial cuts as indicated by the black line in Fig.~\ref{fig:figure8}a.\\
We use this background signal to convert between experimental and numerical units as follows. We perform a numerical simulation of the background oscillation (without additional bion imprint) at $q/h \sim 0$ with similar amplitude in $Q_0$ as in the experimental data. This signal serves as an independent reference for the relevant timescale for spin dynamics.
We similarly extract $f_{b, sim}$ in numerical units. As we work at $q/h\sim0$ both in numerical simulations and in the experiment, we can convert between unit systems via
\begin{equation}
        t_{s,exp} = t_{s,sim} \cdot f_{b, sim}/f_{b, exp}\,,
\end{equation}
where $t_{s} = \hbar / (n |c_1|)$ and we assume constant background density $n$.\\
The error on $t_{s,exp}$ is obtained from the error of $f_{b,exp}$ via error propagation.
To find a corresponding characteristic length scale concerning the topological bion width, we estimate the healing length $\xi_{s} = \hbar/\sqrt{2 M n\lvert c_1\rvert}$ and calculate the experimental scale as
\begin{equation}
    \xi_{s,exp} = \sqrt{\hbar t_{s,exp}/(2 M)} \,.
    \label{eq:equationB3}
\end{equation} 
The error is propagated accordingly and gives rise to the error-bands depicted in Fig.~\ref{fig:figure3} in the main text.\\ This conversion can be done in a similar manner by employing the width of a single sine-Gordon-like soliton which is given by $l = \xi_s / \sqrt{2}$. We follow the procedure described in~\cite{Deller2026} to prepare a sine-Gordon-like soliton in the experiment and extract its width. We then use Eq.~\eqref{eq:equationB3} to calculate the characteristic timescale. Both methods yield the same results within the errors as can be seen in Tab.~\ref{tab:table1}
%==============================================================================
\begin{table}[ht]
    \centering
    \begin{tabular}{|c|c|c|}
        \hline
         Method  & $\xi_{s,exp}\,[\mathrm{\mu m]}$ & $t_{s,exp}\,[\mathrm{ms}]$ \\ \hline
        Background oscillation   & $6.37(9)$ & $111(3)$ \\ \hline
        Soliton width   & $6.43(22)$ & $113(8)$ \\ \hline
    \end{tabular}
        \caption{Characteristic time scale $t_s$ and length scale $\xi_s$ used to map the numerical data to experimental units. These were extracted from the measurements in two independent ways (see text).}
    \label{tab:table1}
\end{table}
%==============================================================================
%==============================================================================
%==============================================================================

%==============================================================================
%==============================================================================
\section{Analytical solution of the Landau-Lifshitz equation}
\label{app:c}
In this section we list the remaining components of the vector that solve the Landau-Lifshitz equation governing the relevant $\mathrm{SU}(2)$ subspace. We also give the derivations for the frequency dependence of the bions with regard to the amplitude $A$ and width $d$.\\
As discussed in the main text the observables $F_x$, $\Qyz$ and $Q_0$,  are obtained from the mean field wave functions $\psi_{m_F}$ of the individual spin components by evaluating $\psi_{m'_F}^{*} o_{m'_F m_F} \psi_{m_F} / n$ with summation over repeated indices and the matrices $o_{m'_F m_F}$ given by
\begin{equation}
\label{app:equationC1}
\begin{gathered}
f_x = \frac{1}{\sqrt{2}}
\begin{pmatrix}
0 & 1 & 0\\
1 & 0 & 1\\
0 & 1 & 0
\end{pmatrix},
\qquad
q_{yz} =
\frac{1}{\sqrt{2}}
\begin{pmatrix}
0 & -i & 0\\
i & 0 & i\\
0 & -i & 0
\end{pmatrix},
\\[0.6em]
q_{0} =
\begin{pmatrix}
-1 & 0 & 0\\
0 & 1 & 0\\
0 & 0 & -1
\end{pmatrix}\,,
\end{gathered}
\end{equation}
and, for completeness,
\begin{align}
\label{app:equationC2}
f_y =
\frac{1}{\sqrt{2}}
\begin{pmatrix}
0 & -i & 0 \\
i & 0 & -i \\
0 & i & 0
\end{pmatrix},
\qquad
f_z =
\begin{pmatrix}
1 & 0 & 0 \\
0 & 0 & 0 \\
0 & 0 & -1
\end{pmatrix}.
\end{align}

%==============================================================================
\subsection{Non-topological bion}
\label{app:c1}
Considering the non-topological bion the components of $\boldsymbol{m} = (F_x, \Qyz,Q_{\mathrm{0}})$ are given by~\cite{Schmitt2026}

\begin{align}
F_x^{\mathrm{nT}}(X,T)
&=
1-\frac{2(1-\Omega)}
{1+\Omega\sinh^2\!\left(
\sqrt{1-\Omega}\,X
\right)}\,,
\label{app:equationC3}
\\
Q_{yz}^{\mathrm{nT}}(X,T)
&=
\frac{2\sqrt{\Omega(1-\Omega)}\cosh\!\left(\sqrt{1-\Omega}X\right)}
{1+\Omega\sinh^2\!\left(\sqrt{1-\Omega}X\right)}
\cos(\Omega T)\,,
\label{app:equationC4}
\\
Q_0^{\mathrm{nT}}(X,T)
&=
\frac{2\sqrt{\Omega(1-\Omega)}\cosh\!\left(\sqrt{1-\Omega}X\right)}
{1+\Omega\sinh^2\!\left(\sqrt{1-\Omega}X\right)}
\sin(\Omega T)\,.
\label{app:equationC5}
\end{align}
In order to compare the experimental and numerical results with the analytical prediction, we introduce the phase amplitude $A$, defined by the angular excursion along the spin-nematic sphere as illustrated in Fig.~\ref{fig:figure2}c, and recover the frequency dependence given by Eq.~\eqref{eq:equation6}. We define the phase amplitude as 
\begin{equation}
    A=\arctan\!\left[\Qyz(0,0)/F_x(0,0)\right] = \arctan\!\frac{2\sqrt{\Omega(1-\Omega)}}{2\Omega-1}\,.
    \label{app:equationC6}
\end{equation}
Since $0<\Omega<1$ we parametrize $\Omega = \cos^2\theta$ with $\theta \in (0,\pi/2)$. We use this to rewrite \eqref{app:equationC6} and using appropriate trigonometric identities we find 
\begin{equation}
    A = 2\theta\,.
\end{equation}
Plugging this back into the definition of $\theta$ we find
\begin{equation}
    \Omega = \cos^2\left(\frac{A}{2}\right)\,.
\end{equation}

%==============================================================================
\subsection{Topological bion}
\label{app:c2}
Considering the topological bion solution the components of $m$ are given by~\cite{Schmitt2026}
\begin{align}
F_x^{\mathrm{T}}(X,T)
&=
1-\frac{2(1+\omega)}
{1+\omega\cosh^2\!\left(
\sqrt{1+\omega}\,X
\right)}\,,
\label{app:equationC9}
\\
Q_{yz}^{\mathrm{T}}(X,T)
&=
\frac{2\sqrt{\omega(1+\omega)}\sinh\!\left(\sqrt{1+\omega}X\right)}
{1+\omega\cosh^2\!\left(\sqrt{1+\omega}X\right)}
\cos(-\omega T)\,,
\label{app:equationC10}
\\
Q_0^{\mathrm{T}}(X,T)
&=
\frac{2\sqrt{\omega(1+\omega)}\sinh\!\left(\sqrt{1+\omega}X\right)}
{1+\omega\cosh^2\!\left(\sqrt{1+\omega}X\right)}
\sin(-\omega T)\,.
\label{app:equationC11}
\end{align}
In order to determine the dependence of $\omega$ given by~\eqref{eq:equation7} we define the width $d$ as the distance between the extrema of the bion in the observable $Q_{yz}$. To derive the positions of the extrema we write the first spatial derivative as 

\begin{align}
\left. \frac{\partial Q_{yz}^{\mathrm{T}}}{\partial X} \right|_{T=0}
&=
\frac{
2\sqrt{\omega}(1+\omega)
\cosh\!\left(\sqrt{1+\omega}X\right)
}
{
\left[
1+\omega\cosh^2\!\left(\sqrt{1+\omega}X\right)
\right]^2
}
\nonumber\\
&\quad\times
\left[
1+\omega-\omega\sinh^2\!\left(\sqrt{1+\omega}X\right)
\right].
\label{app:equationC12}
\end{align}
and solve for $\partial Q_{yz}^{\mathrm{T}}/\partial X = 0$. In turn, \eqref{app:equationC12} simplifies to
\begin{equation}
    1+\omega-\omega\sinh^2\left(\sqrt{1+\omega}X_\pm\right) = 0\,,
\end{equation}
and we find the respective positions as
\begin{equation}
    X^\pm = \pm\frac{1}{\sqrt{1+\omega}}\mathrm{arsinh}\left(\sqrt{\frac{1+\omega}{\omega}}\right)\,.
\end{equation}
The same expression is found for the zero-crossings of $F_x^{\mathrm{T}}$.
Defining the width as $d = x^+-x^-$ and plugging in the definition of $X= (x-x_0)/\ell$, it results as
\begin{align}
    d(\omega) &= \frac{2\ell}{\sqrt{1+\omega}}\mathrm{arsinh}\left(\sqrt{\frac{1+\omega}{\omega}}\right)\\
    &= \frac{\ell}{\sqrt{1+\omega}}\mathrm{arcosh}\left(3+\frac{2}{\omega}\right)\,.
\end{align}

%==============================================================================
%==============================================================================
\section{Numerical simulations of the spin-1 system}
\label{app:d}
We numerically compute the mean-field spinor wave function $\psi = (\psi_{1}, \psi_{0}, \psi_{-1})^{T}$ by solving the three-component Gross-Pitaevskii equation (GPE),
\begin{align}
i\hbar \frac{\partial \psi(x,t)}{\partial t}
=&
\left[
-\frac{\hbar^2}{2M}\frac{\partial^2}{\partial x^2}
% + V_{\mathrm{ex}}(x)
+ q f_z^2 \right. \\
&\left.\ +\  \frac{c_0}{2}\left|\psi(x,t)\right|^2
+ \frac{c_1}{2}\,\mathbf{F}(x,t)\cdot\mathbf{f}
\right]\psi(x,t)\,. \nonumber
\end{align}
Here, $\mathbf{f}$ denotes the vector $(f_x,f_y,f_z)$ that contains the three spin matrices, which are defined in Eqs.~\eqref{app:equationC1} and \eqref{app:equationC2}.

The spin density is then given by 
\begin{align}
    \mathbf{F}
=
\begin{pmatrix}
F_x \\
F_y \\
F_z
\end{pmatrix}
=
\frac{1}{\sqrt{2}}
\begin{pmatrix}
\psi_0^{*}\left(\psi_{+1}+\psi_{-1}\right)+\mathrm{c.c.} \\
i\,\psi_0^{*}\left(\psi_{+1}-\psi_{-1}\right)+\mathrm{c.c.} \\
\sqrt{2}\left(|\psi_{+1}|^2-|\psi_{-1}|^2\right)
\end{pmatrix}. 
\end{align}
The density and spin couplings are given by $c_0$ and $c_1$, respectively. In this work, we only consider vanishing quadratic Zeeman shifts, $q/h=0$. This is a boundary of the ground-state phase diagram of the spin-1 Hamiltonian \cite{SCHMIEDGS}. We initialize the system in a transversally elongated ground state,
\begin{align}
    \psi_{\mathrm{EP}}
=
\frac{\sqrt{n}}{2}
\begin{pmatrix}
1 \\
\sqrt{2} \\
1
\end{pmatrix}.
\end{align}

%==============================================================================
\subsection{Initial conditions}
We prepare the initial conditions for the numerical analysis that corroborates the experimental results as depicted in Fig.~\ref{fig:figure1}, which are anticipated to give stable objects for the spin-1 gas by imprinting profiles on the spinor phase $\phis$. We prepare a ground state of $1.5\times10^4$ atoms on a discrete periodic grid with a physical length of $220\,\mathrm{\mu m}$ and $4096$ grid points. This resolves the characteristic length scale, the spin healing length, $\xi_s$ of the system well, as it corresponds to $\sim 40.8$ grid points. \\
For the numerical simulations presented in this publication we recreate the experimental imprint scheme of advancing the spinor phase by a sum of Gaussians, which correspond to the Gaussian beams applied to the atom cloud in the experiment.
\end{appendix}
\twocolumngrid
\bibliographystyle{apsrev4-1}

\bibliography{Bibliography}% Produces the bibliography via BibTeX.
%==============================================================================
%==============================================================================

\end{document}